# Developing and Evaluating Quilts for the Depiction of Large Layered Graphs

Juhee Bae and Ben Watson, *Member, IEEE*

**Abstract**—Traditional layered graph depictions such as flow charts are in wide use. Yet as graphs grow more complex, these depictions can become difficult to understand. Quilts are matrix-based depictions for layered graphs designed to address this problem. In this research, we first improve Quilts by developing three design alternatives, and then compare the best of these alternatives to better-known node-link and matrix depictions. A primary weakness in Quilts is their depiction of skip links, links that do not simply connect to a succeeding layer. Therefore in our first study, we compare Quilts using color-only, text-only, and mixed (color and text) skip link depictions, finding that path finding with the color-only depiction is significantly slower and less accurate, and that in certain cases, the mixed depiction offers an advantage over the text-only depiction. In our second study, we compare Quilts using the mixed depiction to node-link diagrams and centered matrices. Overall results show that users can find paths through graphs significantly faster with Quilts (46.6 secs) than with node-link (58.3 secs) or matrix (71.2 secs) diagrams. This speed advantage is still greater in large graphs (e.g. in 200 node graphs, 55.4 secs vs. 71.1 secs for node-link and 84.2 secs for matrix depictions).

## 1 Introduction

There are numerous ways to depict layered graphs, including the well known node-link diagrams, lesser known matrices, and most recently, Quilts. Node-link diagrams (Figure 1(a)) are easy to read and understand but as graphs become more complex (with more nodes and links), they suffer from link overlap and occlusion [1]. Matrices [2] assign a unique spatial position to each possible link, avoiding the occlusion problem, but graph complexity can make them quite large. Quilts [16, 3], which use an alternative matrix-based depiction for layered graphs, are more compact. However, skip links (which do not connect to a succeeding layer) can be hard to follow in Quilts, since finding the link destination requires color matching.

In this paper, we first improve Quilts by developing two new skip link depictions, and then identifying the best of the three alternatives in an experimental evaluation using a path-finding task (a known weakness of matrix depictions). Next, we compare our improved Quilt depiction with the node-link and centered matrix depictions, learning that Quilts enable viewers to find paths more quickly and accurately than do node-link diagrams and matrices. We also obtain interesting results indicating the relationship of several layered graph characteristics to path finding performance.

## 2 Related Work

Node-link diagrams [1, 11, 15] (Figure 1(a)) are widely used and are easy to read when graphs are simple. However, Ghoniem et al. showed that the legibility of these diagrams drops steeply when they contain 50 nodes or more [8]. On the other hand, matrices [2] remain much more legible as the number of nodes in graphs grows. Ghoniem et al. [8] showed that among seven graph reading tasks, node-link diagrams proved to be better only for path finding, and this advantage disappeared when graphs grew to contain 100 nodes. Keller et al. [12] also compared node-link and matrix based depictions by performing two experiments using several different reading tasks. Their first experiment, in which participants counted a node's incoming or outgoing links, found that increases in size and density significantly reduced performance. Their second experiment had participants perform several different tasks, and used data set drawn from two applications,

- *Juhee Bae is with NC State University, E-mail: jbae3@ncsu.edu.*
- *Ben Watson is with NC State University, E-mail: bwatson@ncsu.edu.*

both containing less than 50 nodes and 150 links. Results confirmed Ghoniem et al.'s conclusions: for small, sparse graphs, node-link diagrams are easier to read; while for more complex graphs, matrix depictions enable better understanding. Both of Ghoniem et al. and Keller et al. examined only unlayered graphs.

Researchers have developed several improvements of the basic matrix representation, attempting to preserve its advantages while addressing its weaknesses. Shen and Ma [13] proposed *centered matrices* (our own term), which move nodes from the beginning of rows and columns to the matrix diagonal. This simplifies visualized paths (and one would assume, path following) significantly. NodeTrix [10] proposes using matrices for dense subgraphs, and connecting these subgraphs with node-link diagrams. In a case study using information visualization citation data, this combination supported both learning global graph structure and understanding of detailed subgraphs. MatLink [9] supplements matrix depictions with links drawn along row and column headings, permitting path visualization without occluding the matrix itself. In a comparison with traditional matrix and node-link diagrams using applied data with fewer than 100 nodes and less than 30% link density, users performed several different reading tasks. MatLink always supported performance that was as good as or better than matrix depictions. Like the matrices in Ghoniem et al. [8], MatLink supported good performance on simpler, lower level tasks. On the other hand, those tasks were difficult to perform with node-link diagrams because of variations in link density. For path related tasks, MatLink was at least as good as node-link diagrams, and in fact outperformed node-link diagrams for finding shortest paths, addressing the major shortcoming of matrices identified by Ghoniem et al. For tasks identifying global link structure, node-link diagrams still offered some advantages.

Layered graphs, such as flow charts and genealogy diagrams, are typically represented in node-link diagrams by arranging the nodes in each layer into horizontal rows. There has been extensive research into finding the optimal layout within these constraints, using criteria such as link crossings and link complexity [1, 15, 5]. In matrices, layered graphs can easily be represented by arranging nodes into groups corresponding to layers, and highlighting them.

Quilts [16] chain matrices representing proper links (links that do not skip, moving from layer to succeeding layer) in a cascading series connected by layers. Layers are formed of rows and columns of cells, with each layer having a unique chromaticity (combination of hue and saturation). Figures 1(a) and 1(b) show a simple Quilt and a corresponding node-link depiction of the same graph. Identical graph nodes have the same color in both figures. In Quilts, rectangular, colored cells represent nodes, while black and colored circles represent proper and skip links. To follow a skip link, viewers must move from the node at the head of its row or column to the node that matches the color of the skip link. For example in Figure 1(b), the dark green skip link represents a link between the dark red and the dark green nodes.

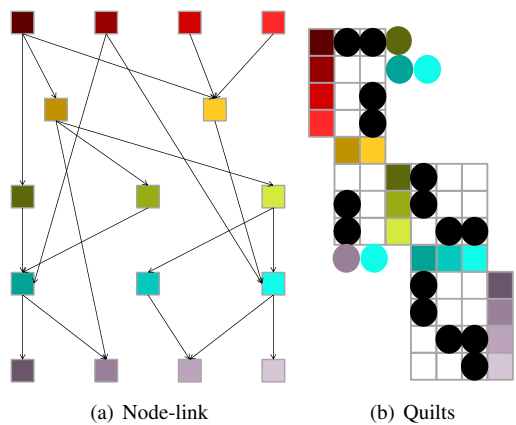

(a) Node-link    (b) Quilts

Fig. 1. Node-link and Quilt depictions of the same graph.

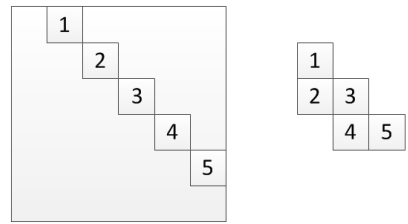

Fig. 2. Comparing the layout of proper links in matrices and Quilts.

Essentially, Quilts extract the submatrices that represent proper links in traditional matrix depictions, and rearrange them in a spatially more efficient manner. In matrix depictions, these submatrices are above and to the right of the matrix diagonal (Figure 2, left). Because each node in a matrix depiction occupies a unique range of *both* horizontal and vertical space, each of the matrix's proper link submatrices do the same. On the other hand, submatrices in Quilts (Figure 2, right) occupy a unique range of *either* horizontal or vertical space, giving Quilts roughly half the height and width of the corresponding matrices, and roughly a quarter the area.

The tradeoff for the spatial efficiency of Quilts is that the destination nodes of skip links are harder to see, because they can no longer be represented by the skip link's spatial position, as in matrices. Instead, color or some other display element must be used. In graphs with many nodes, matching color to follow skip links can become difficult.

Recently, Bezerianos et al. [3] applied Quilts to genealogy. They were able to gain all the advantages of Quilt compactness without introducing any new difficulty in following skip links. Their application used a bipartite graph, with layers representing individuals alternating with layers representing families. Because individual nodes could only link to family nodes and vice versa, position could be used to depict skip link destinations. It remains to be seen how many more applications will also permit a similar use of bipartite graphs.

In sum, Quilts are more compact than matrices, but generally use a non-spatial depiction of skip link destinations. Quilts are not obviously smaller than node-link diagrams, but do not suffer from link crossing problems. Because maximizing compactness and minimizing crossings are particularly important when graphs are complex, it is reasonable to conjecture that Quilts may offer advantages when layered graphs are complex. We decided to test this conjecture experimentally.

## 3 EXPERIMENT 1

We begin our experimental work by designing two new skip link depictions for Quilts, illustrated in Figure 3. At left, the existing color-only depiction distinguishes each layer with different chromaticity and each node within a layer with a different brightness. At right, the text-only depiction assigns a letter to each layer, and a number to each node within a layer. In the center, the mixed depiction assigns a color to each layer, and a number to each node within a layer. For this depiction, we could have double-coded the within-layer node position using both color *and* numbers, but we chose to avoid any ambiguity that would result by continuing to support color matching. We compared these three techniques for depicting skip links experimentally. We expected to learn that the mixed and text-based depictions of skip links were more legible, since they did not rely solely on color matching.

We chose path finding as our evaluation task because it is a high level activity involving basic tasks including node finding, link finding, finding the most connected (highest degree) node, and finding neighbors. Moreover, as we have already described, previous research (e.g. Ghoniem et al., [8]) identifies path finding as a weakness of matrix depictions. We reasoned that a successful demonstration of an advantage for matrix-based Quilts at what prevailing opinion holds is matrices' weakest point would be particularly powerful.

### 3.1 Methods
#### 3.1.1 Design

We used a five-factor (3 *depictions* x 3 *nodes* x 2 *links* x 2 *skips* x 3 *layers*) within-subjects design. *Depiction* of skip links had three levels: color, text, and mixed. The number of *nodes* also had three levels: 50, 100 and 200. The density of proper *links* varied between 25% and 50% of all proper links possible. For example, given 400 cells in a Quilt's matrices, a density of 25% would result in 100 proper links, 50% in 200 proper links. The density of *skips* varied between 25% and 50% of the number of proper links (were skip density much higher, the graph would not be a good candidate for layering). Finally, the number of *layers* varied across 5, 10 and 15.

#### 3.1.2 Participants
18 college students (11 male, 7 female, aged from 19 to 38) participated in the experiment. All had normal or corrected-normal vision and passed a color-blindness test.

#### 3.1.3 Apparatus
The experiment was performed on a Dell workstation with NVIDIA GeForce 7950 GX2, an Intel Core Duo CPU 2.4 GHz processor, 4GB of RAM, Windows 7 OS, and a 1920 x 1200 pixel Dell 24" monitor. The graphs were displayed in a full screen mode. Participants sat on an office chair in any fashion they found comfortable.

#### 3.1.4 Stimuli
With 36 different graph treatments (3x2x2x3) and 3 different depictions, we required 108 graphs per session. We randomly generated directed, layered graphs as specified by the experimental variables above in two *generate* and *test* stages. In the *generate* stage, a Python script given a experimentally specified number of *nodes*, *layers*, *links* and *skips* generated a graph with the passed number of nodes, and randomly assigned a layer number between 1 and *layers* to each of them. The script then calculated the total number of proper links possible, and with *links* and *skips*, the total number of proper and skip links needed. The script randomly located the needed number of proper links in the Quilt's matrices, and the needed number of skip links between other (non-proper) node pairs. Note that skip links could connect to nodes on the same layer, or a higher or lower numbered layer, and so often moved "upward" in the three depictions.

In the second *test* stage, a Python script checked that the graphs generated in the first stage met several other experimental constraints. If a graph did not meet the constraints, it generated another graph with the same experimental variables and tested it, continuing until it had all the graphs required. There were four experimental constraints. First, the graph had to contain at least one "good" path between a randomly chosen source node in the first layer, and a randomly chosen destination node in the last layer. Second, since we were evaluating skip link depictions, any "good" path from source to destination had to contain at least one skip link, and therefore could not contain more links than the number of layers minus two. Third, to enforce a minimum difficulty, any "good" path had to contain at least three links. Fourth, the Quilt depiction of the graph had to fit on our display (we rejected only six graphs this way).

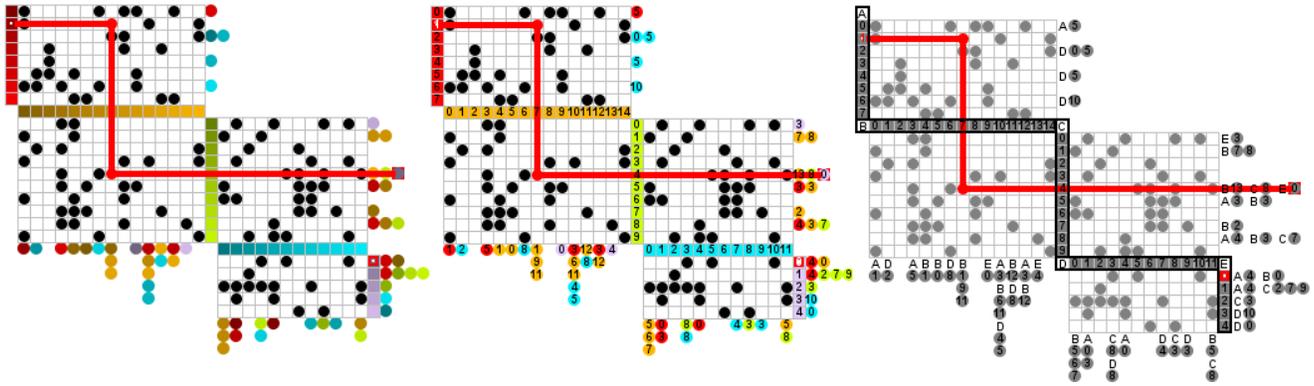

Fig. 3. A 50 node, 25% link density, 50% skip link density, and 5 layered graph, shown using the color-only, mixed, and text-only depictions. The red highlight indicates a successful path from source to destination, while the source and destination nodes are highlighted with white dots (color-only), white letters (mixed) or red letters (text).

We wrote custom software to display Quilts, and to support and record path-finding interaction with them. Participants traced a path on the graph itself by clicking on nodes or links with the mouse. If the node or link could be reached from the last clicked node or link, we highlighted it in red. There was no mouseover highlighting. To backtrack, we supported clicking on previously highlighted nodes or links. Path-building could proceed downward from the source node, upward from the destination node, or both. We detected successful completion of a path automatically by verifying that it connected the source and destination nodes, and met the length constraints of a "good" path. If a path was not completed within four minutes, the path-finding task was halted. In either case, we displayed the time used. The next graph would not appear until the participant pressed a key, offering the participant an opportunity to take a break after every displayed graph.

Brightness in colored Quilts varied per layer from maximum to a non-black floor and did not vary in mixed Quilts. Initially, we varied chromaticity in colored Quilts algorithmically by distributing samples evenly in RGB color space. However in piloting, these results proved unsatisfactory, likely due to perceptual nonlinearities. Rather than implementing a complex color mapping optimization algorithm, we defined a color mapping manually, and tuned it in pilot studies to maximize visual differences between 5, 10, and 15 layers. The final result defined a different chromaticity for each layer.

### 3.1.5 Procedure

To avoid confounding our experimental variables with learning and fatigue, we varied the order of our experimental treatments. We used complete counterbalancing of depiction. Within each depiction, we sampled the orders of the remaining treatments with a Latin Square.

We obtained informed consent from the participants, and gave them written instructions explaining Quilts, the various skip link depictions, and their path-finding task. We instructed participants to complete each trial as quickly as possible, even if they may have to undo mistakes. Participants first practiced the path-finding task on paper, then on the interactive system. The experiment began when participants felt ready. They performed two sessions for a total of 216 graphs, breaking for at least two hours between sessions. They were paid $20 for their effort. As dependent measures, we recorded time to trace the path, and whether or not the path was indeed found ("accuracy").

### 3.2 Hypotheses

As noted above, we expected that the color-only depiction would be least legible as measured by both time and accuracy. This would be particularly true when the number of nodes was large, making color matching harder. We also expected that the mixed depiction would be more legible than the text-only depiction since it enabled finding layers more quickly, using color rather than textual matching. This reliance on color matching would make the mixed depiction's advantage most apparent when the number of nodes was small.

As to effects not directly related to skip link depiction, we expected that more nodes would make following paths harder, since it increases graph complexity and the participant's search space. More links make finding paths easier, as it did for matrices in Ghoniem et al. [8]. Variations in both skip links and layers introduced complex tradeoffs (e.g., skip links can shorten paths, but are hard to follow; more layers simplifies color matching, but also lengthens proper link paths), so we did not hazard any predictions about their effects.

### 3.3 Results

We analyzed our results with a five-factor repeated-measures analysis of variance (ANOVA). We discuss the main significant effects, and the main significant two-way interactions. We detail significant effects on the time measure in Table 1, and on the accuracy measure in Table 2. All the pairwise comparisons were evaluated with contrasts.

*Depictions* had a significant effect on the time measure, with means depicted in Figure 4. Pairwise comparisons showed that participants found paths more slowly with the color depiction than with the mixed and text depictions. Depiction also had a significant effect on accuracy, with means of 93%, 98% and 97% for the color, mixed and text depictions respectively. Pairwise comparisons showed that participants were less likely to find paths with the color depiction than with the mixed and text depictions.

All of the remaining variables also had significant main effects. *Nodes* had a significant effect on the time measure, with path finding time increasing as nodes increased, also shown in Figure 4. Pairwise comparisons showed that each increase in the number of nodes made path-finding slower. *Links* had a significant effect on time, with means of 60.83 and 44.65 seconds for link densities of 25% and 50% respectively. Links also had a significant effect on accuracy, with means of 95% and 97% for the 25% and 50% link densities. Participants were faster and more accurate when proper links were denser. *Skips* had a significant effect on time, with means of 50.7 and 54.77 seconds for skip link densities of 25% and 50% respectively. Participants found paths faster when there were fewer skip links. Finally, *layers* had a significant effect on time, with path finding time decreasing as the number of layers increased, as shown in Figure 4. Pairwise comparisons showed that participants found paths more quickly with each increase in the number of layers. *Layers* also had a significant effect on accuracy, with means of 94%, 97% and 98% for 5, 10 and 15 layers. Pairwise comparisons showed that participants found paths more reliably with 10 and 15 layer graphs than with 5 layer graphs.

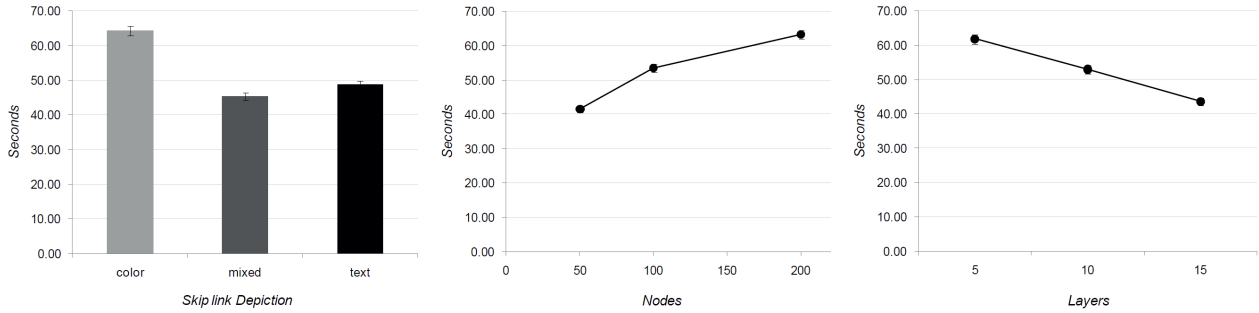

Fig. 4. Path finding times in Experiment 1 as affected by depiction, nodes and layers. Bars show standard error.

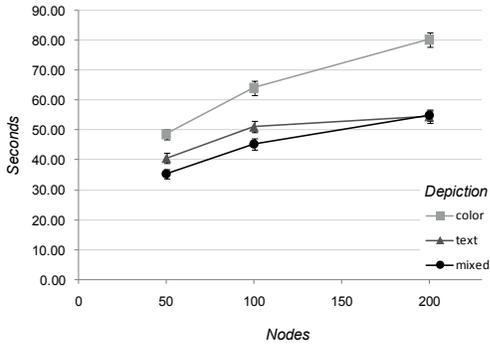

(a)

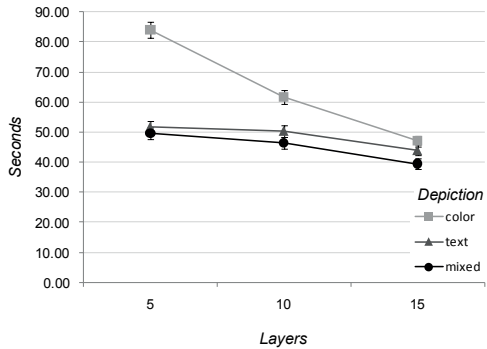

(b)

Fig. 5. Path finding times in Experiment 1, as affected by depiction vs. nodes, and depiction vs. layers. Bars show standard error.

The interaction between *depictions* and *nodes* was significant by both the time (Table 1) and accuracy (Table 2) measures. We show the two-way time means in Figure 5(a). As the number of nodes increased, participants using the color skip link depiction took longer to find paths, and were less successful at finding paths, than those using the mixed and text depictions. Moreover, two-way comparisons show that at 50 and 100 nodes, performance with the mixed depiction was faster than text depiction, but at 200 nodes, performance with the two depictions did not differ. The interaction between *depiction* and *layers* was also significant by the time (Table 1) and accuracy (Table 2) measures. We show the two-way time means in Figure 5(b). As the number of layers decreased, participants using the color skip link depiction took longer to find paths, and were less successful in finding paths, than those using the mixed and text depictions. In other words, there was drastic improvement in time and accuracy with color depiction as the number of layers increased.

| independent variable | ANOVA of time |
|---|---|
| depictions | $F(2,34)=25.492, p<.00005$ |
| nodes | $F(2,34)=67.118, p<.00005$ |
| links | $F(1,17)=83.238, p<.00005$ |
| skips | $F(1,17)=6.461, p<.05$ |
| layers | $F(2,34)=25.586, p<.00005$ |
| depictions*nodes | $F(4,68)=2.801, p<.05$ |
| depictions*layers | $F(4,68)=10.437, p<.00005$ |

Table 1. Significant main effects on path finding time in Experiment 1, as well as significant two-way interactions with depiction.

| independent variable | ANOVA of accuracy |
|---|---|
| depictions | $F(2,34)=12.218, p<0.0005$ |
| links | $F(1,17)=14.825, p=.001$ |
| layers | $F(2,34)=17.335, p<.005$ |
| depictions*nodes | $F(4,68)=4.759, p=.002$ |
| depictions*layers | $F(4,68)=8.91, p<.0005$ |

Table 2. Significant main effects on path finding accuracy in Experiment 1, as well as significant two-way interactions with depiction.

### 3.4 Discussion

The results of our first experiment clearly show that Quilts using the color-only skip link depiction do not preserve the legibility of large graphs well. Moreover, pairwise comparisons show that Quilts using the mixed depiction have a slight advantage over the text-only depiction. It may be that color supports quicker identification of layers than text, since it accesses a lower perceptual level than text.

The main effects of most of our other variables are easily understood. Increasing nodes makes the path-finding task more difficult and complex. Increasing links eases the path-finding task by increasing the number of paths. It is a bit surprising that more skip links increased the time to find paths, until one recalls that in our design, use of a skip link was required, and having more skip links generally made a useful skip link harder to find. Most intriguing, however, is that more layers dramatically improves path finding time. This is likely due to a number of factors. First, with our constraints on path length for the first experiment, a 5-layer graph could only have "good" paths of length three: shorter and longer paths were rejected. With 10- and 15-layer graphs, a larger range of path lengths were accepted. In addition, as the number of layers decreased, the average number of nodes in each layer increased, increasing the difficulty of color matching. And more generally, more layers makes skip links more helpful: they can skip closer to the path's end.

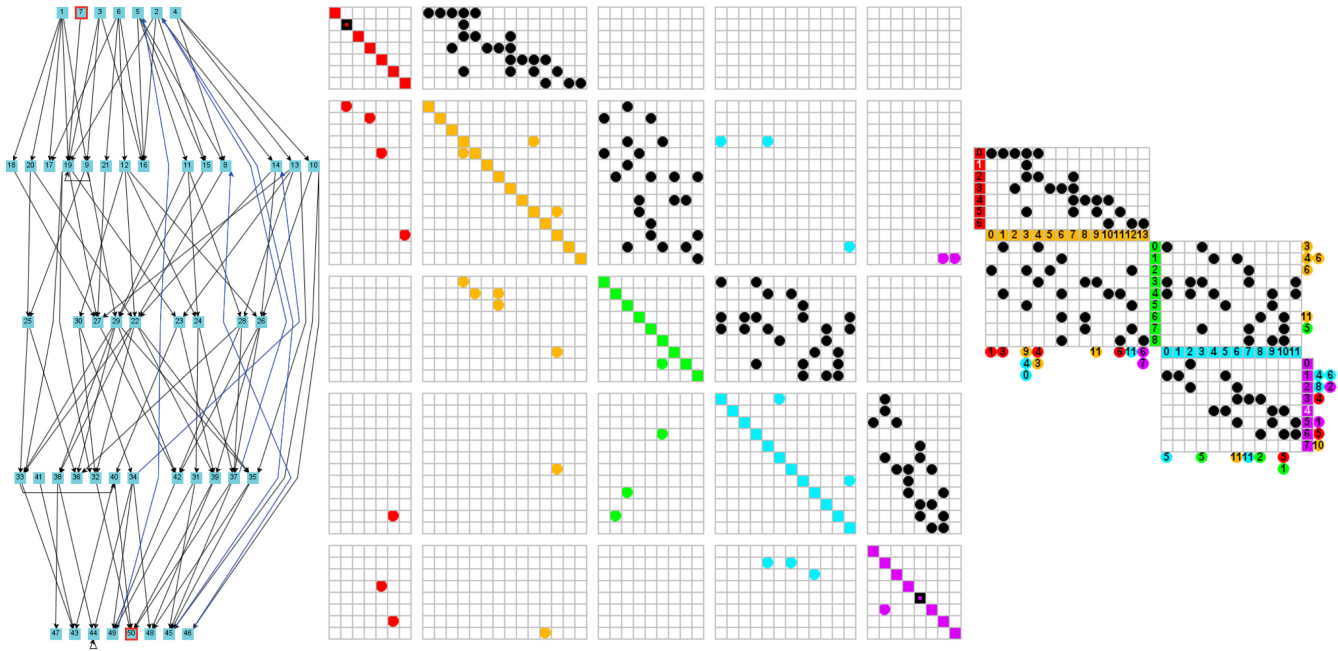

Fig. 6. Example of a 50 node, quarter link density, quarter skip link density, and 5 layered graph. The left most depicts a node-link diagram, middle shows a centered matrix, and right most is Quilts. Red and black boxes and white numbers respectively indicate the source and destination node. Blue links in a node-link diagram indicate backward links. The node-link diagram is relatively much larger in experimental display.

## 4 EXPERIMENT 2

With the knowledge that Quilts using mixed skip link depictions are the easiest to read, we wanted to compare them to two better known alternative depictions: node-link diagrams and centered matrices, illustrated in Figure 6. We chose to use centered rather than uncentered matrices (with nodes at row and column headings) because we expected that they would provide better support for path finding. We anticipated that Quilts would support faster and more accurate path finding than these alternatives.

### 4.1 Methods

Our second experiment has many similarities to our first. Below, we note only the differences between the second and first experiments.

#### 4.1.1 Design

We again used a five-factor (3 *depictions* x 3 *nodes* x 2 *links* x 2 *skips* x 2 *layers*) within-subjects design. In this experiment, however, the three levels of *depictions* were: node-link diagram, centered matrix, and Quilts. The density of *skips* varied between 0% and 25% of the number of proper links. This change allowed us to examine the effect of having skip links at all. Because the overall trend of *layers* was linear in experiment 1, we eliminated the middle level, leaving only graphs with 5 and 15 layers.

#### 4.1.2 Participants

24 college students (17 male, 7 female, aged from 20 to 56) participated in the second experiment. All had normal or corrected-normal vision and passed a color-blindness test.

#### 4.1.3 Apparatus

The experiment was performed on a Dell workstation with NVIDIA GeForce 7950 GX2, an Intel Core Duo CPU 2.4 GHz processor, 4GB of RAM, Windows 7 OS, and a 2560 x 1600 pixel Apple 30" monitor. The graphs were displayed in a full screen mode. Participants sat on an office chair in any fashion they found comfortable.

#### 4.1.4 Stimuli

We reused the 25% link density graphs from Experiment 1 and removed skip links from the 50% link density graphs for non skip link graphs. We retained the constraints requiring at least one "good" path, and a minimum path length in all good paths of at least three links. Because our experiment was no longer focused on skip links, we allowed the maximum path length to rise to 1.5 times of the number of layers. To fit the node-link diagram and matrix depictions onto our display, we were forced to use the 2560 x 1600 pixel monitor. With 24 different graph treatments (3x2x2x2) and 3 different graph depictions, we showed 72 graphs to the participants.

To lay out node-link diagrams, we used Stallman et al.'s minimal crossing edge algorithm with a barycentric heuristic [14]. Because experimental control required us to use a fixed number of layers, and to treat each node's layer as a fixed input rather than a variable determined by a layout algorithm, we were not able to use techniques from many of the latest layout algorithms, such as those used by the GraphViz system [7]. The first layer of node-link diagrams was always located at the top of the display, and the last always at the bottom. Node size and intra-layer node separation were also unchanging, with any skip links passing through a layer requiring less separation between one another. Inter-layer separation was constant within each graph, but varied across graphs, with graphs with five layers having a larger separation than graphs with 15 layers.

We scaled centered matrices so that graphs with 200 nodes filled the vertical extent of the display. To avoid any reading and interaction difficulty caused by widely varying cell sizes, we depicted smaller graphs using the same matrix cell size as larger graphs. As a result, these smaller graphs did not fill the vertical extent of the display. We also scaled Quilts so that the graphs with the largest vertical extent in the Quilts depiction filled the vertical extent of the display, and with the same reasoning used for matrices, used an identical cell size across all graphs, whether or not the resulting depiction filled the display. In fact, due to random asymmetries in horizontal vs. vertical distribution of nodes across the Quilts depiction, a few Quilts had more height than width, and the cell size in matrices and Quilts was identical.

We improved our custom software from Experiment 1 to display and support interaction with node-link diagrams and centered matrices. Since clicking links is problematic on dense node-link diagrams, we reminded participants that clicking on nodes was always an option.

### 4.1.5 Procedure

We again avoided confounding our experimental variables with learning and fatigue by varying the order of our experimental treatments. We used complete counterbalancing of depiction, and within each depiction, a Latin Square.

We obtained informed consent from the participants, then asked them to read instructions for the experiment as a whole. We then verbally explained each of the graph depictions and the path finding task to the participants, using paper examples for illustration. We instructed participants to complete their tasks as quickly as possible, even though they might have to undo mistakes. All the participants except one were familiar with node-link diagrams, many were not familiar with centered matrices, and none with Quilts. The time dedicated to explaining each depiction was in general proportional to participant familiarity with the depiction (i.e. the largest amount of time was dedicated to Quilts). Our explanations halted when the participants stated that they understood the depictions and their task. Participants then completed several practice trials: four on paper for each depiction, and three on our interactive system for each depiction. It took about thirty minutes for each participant to complete these practice trials. Participants then began experimental trials, finding paths in 72 graphs in only one session. They were paid $10 for their effort.

### 4.2 Hypotheses

We anticipated that Quilts would provide the best support for path finding, with their compactness and lack of link crossings. Despite their inefficient use of display space, we thought that matrices would provide the next best level of path finding support – although in the research by Ghoniem et al. [8], they were not better overall, trends indicated that matrices would provide better path finding support for large graphs. With their crossings, we assumed that node-link diagrams would perform the worst. We also anticipated that as graph complexity increased as a function of the number of nodes and links, the advantage of using Quilts would improve.

Based on previous research and our experience from Experiment 1, we anticipated that graph legibility would decrease as the number of nodes increased, and increase with the number of links. We did not hazard any predictions about the effects of skip links and layers: we found no precedent in the literature for such effects in matrices and node-link diagrams, and for Quilts, changes in our experimental conditions (use of skips not required, skips not always present) made Experiment 1's implications in our new experimental setting unclear.

### 4.3 Results

We analyzed our results with a five-factor repeated-measures ANOVA. We discuss all significant main, two-way and three-way effects. We detail those effects in Tables 3 and 4. All the pairwise comparisons were made with contrasts.

*Depictions* had a significant effect on the time measure (Table 3), with means depicted in Figure 7(a), and on the accuracy measure (Table 4), with means of 99% (Quilts), 95% (matrices) and 92% (node-link). Pairwise comparisons showed that all means were significantly different from one another. Quilts supported performance that was 35% faster than matrices and 17% faster than node-link depictions.

All of the remaining variables also had significant main effects. *Nodes* had a significant effect on both the time and accuracy measures. Figure 7(b) shows the time means, with the increase from 50 to 100 nodes slowing path finding performance quite a bit, but the increase from 100 to 200 harming performance only slightly. Accuracy dropped from 97% at 50 nodes to 93% at 100 nodes, and then increased back up to 95% at 200 nodes. *Links* had a significant effect on time, with means of 68.34 and 49.01 seconds for link densities of 25% and 50% respectively. Links also had a significant effect on accuracy, with means of 93% and 97% for the 25% and 50% link densities. As in Experiment 1, participants were faster and more accurate when they had greater proper link density. *Skips* had a significant effect on time, with means of 55.11 and 62.23 seconds for skip link densities of 0% and 25% respectively. Participants found paths more quickly when there were no skip links. Finally, *Layers* had a significant effect on time, with means of 53.17 and 64.17 seconds for 5 and 15 layers respectively. In contrast to Experiment 1, path finding times *increased* as the number of layers increased.

| independent variable | ANOVA of time |
| --- | --- |
| *depictions* | F(2,46)=23.541, p<.00005 |
| *nodes* | F(2,46)=62.549, p<.00005 |
| *links* | F(1,23)=32.493, p<.00005 |
| *skips* | F(1,23)=8.225, p=.009 |
| *layers* | F(1,23)=12.219, p=.002 |
| *depictions*links* | F(2,46)=6.739, p=.003 |
| *depictions*layers* | F(2,46)=30.849, p<.00005 |
| *nodes*links* | F(2,46)=15.589, p<.00005 |
| *nodes*layers* | F(2,46)=8.417, p=.001 |
| *links*layers* | F(1,23)=5.483, p<.05 |
| *depictions*nodes*layers* | F(4,92)=3.685, p=.008 |
| *nodes*links*layers* | F(2,46)=5.796, p=.006 |

Table 3. Significant effects on path finding time in Experiment 2.

| independent variable | ANOVA of accuracy |
| --- | --- |
| *depictions* | F(2,46)=12.5, p<0.0005 |
| *nodes* | F(2,46)=5.667, p=.006 |
| *links* | F(1,23)=8.529, p=.008 |
| *depictions*layers* | F(2,46)=13.048, p<.0005 |
| *nodes*links* | F(2,46)=7.196, p<.0005 |
| *nodes*links*layers* | F(2,46)=3.513, p=.038 |

Table 4. Significant effects on path finding accuracy in Experiment 2.

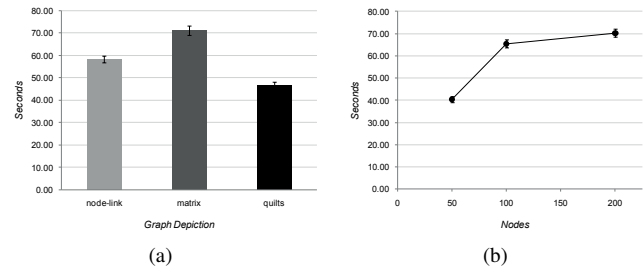

(a)  (b)

Fig. 7. The main effects of depiction and nodes on path finding time in Experiment 2. Bars indicate standard error.

The two-way interaction between *depictions* and *links* was significant for the time measure. We show the two-way time means in Figure 8(a). As link density increased, path finding times dropped much more quickly for matrices than for node-link diagrams or Quilts. The interaction between *depiction* and *layers* was significant by both the time and accuracy measures. We show the two-way time means in Figure 8(b). Corresponding two-way means for accuracy, as layers increased from 5 to 15, were 89% and 95% for node-link diagrams, 97% and 93% for matrices, and 100% and 98% for Quilts. As the number of layers increased, participants using the node-link depiction found paths more quickly and more often, while those using the matrix and Quilts depictions found them less quickly and less often.

There was no significant interaction between *depictions* and *nodes*, meaning that the relative effect of increasing the number of nodes did not change with depiction. Nevertheless all the two-way means were significantly different from one another, and their absolute differences were larger at high node counts. For example, at 50 nodes, path finding times were 33.8 seconds for Quilts, 40 for node-link, and 47.4 for matrices. At 100 nodes, times were 50.5, 63.8, and 81.8 seconds. At 200 nodes, times were 55.4, 71.1, and 84.2 seconds.

The two-way interaction between *nodes* and *links* was significant by both the time and accuracy measures. More links in 100-node graphs improved performance quite a bit, while while in 50- or 200-node graphs, the improvement was smaller. The interaction between *nodes* and *layers* was significant by the time measure. More layers harmed performance in 50- and 100-node graphs, but not in 200-node graphs. The interaction between *links* and *layers* was significant by the time measure. In sparsely linked graphs, increasing layers harmed performance more than it did in densely linked graphs.

The three-way interaction between *depictions*, *nodes* and *layers* was significant by the time measure. Recall that increasing layers when using node-link diagrams improved path finding performance. But increasing layers had little effect when the number of nodes was 100. Recall also that increasing layers when using matrices or Quilts harmed performance. But increasing layers had little effect when the number of nodes was 200. The interaction between *nodes*, *links*, and *layers* was significant on both the time and accuracy measures. Recall that increasing link density when graphs had 100 nodes resulted in a large improvement in path finding performance. But when graphs had only five layers, increasing link density had little effect. In contrast, when graphs had 50 or 200 nodes, varying the number of layers did not fundamentally change the effect of link density on performance.

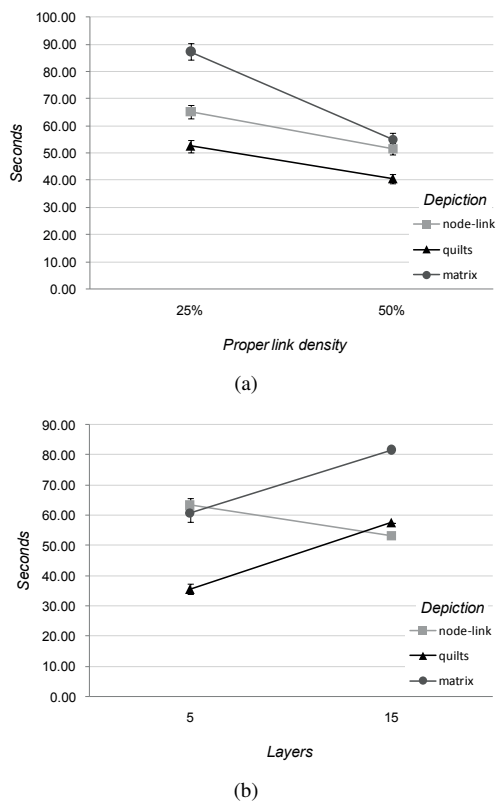

Fig. 8. Path finding times in Experiment 2, as affected by depiction vs. links, and depiction vs. layers. Bars show standard error.

### 4.4 Discussion

The results of our second experiment clearly show that the Quilts depiction supports path finding in complex layered graphs better than node-link diagrams and matrices, even when the number of nodes in those graphs is large. In a small surprise, we also find that path finding with matrices, even in graphs with a large number of nodes, is more difficult than in node-link diagrams. As we noted when discussing our hypotheses, previous research in Ghoniem et al. [8] showed that although matrices were worse for path-finding overall, they were just as effective for graphs containing 100 nodes, a trend we presumed would continue to improve for graphs containing 200 nodes.

As in the first experiment, increasing the number of *nodes* (without varying link density) makes path finding more difficult. However, this effect seems to reach a threshold: increasing the number of nodes in an already complex graph has less effect than an increase in a relatively simple graph. Perhaps the difficulty of tracing paths through a larger graph is mitigated by a corresponding increase in paths to the destination. Increasing *link* density eases path-finding, likely by increasing the number of paths. This was especially true when using matrix depictions, which support path finding particularly poorly.

As to features unique to layered graphs, we learn that having *skips* generally slows path finding performance. Apparently, the advantage of being able to shorten the length of the path followed was outweighed by the cost of searching for those shortcuts. The absence of any *skips\*depictions* interaction indicates that this is true whether the depiction is Quilts, matrices, or node-link diagrams. Increasing the number of *layers* makes path finding harder when graphs are depicted with matrices or Quilts; likely graphs with more layers force viewers to trace longer paths. However when graphs are depicted using node-link diagrams, increasing the number of layers improves path finding performance. Since using more layers reduces the number of nodes on each layer, this brings nodes closer to the display center, making links more vertical. The well known advantage of vertical links and paths in node-link diagrams apparently compensates for the increased length of those paths.

It is worth discussing an apparent contradiction between Experiment 1, which found that increasing layers improved path finding performance; and Experiment 2, which found that increasing layers when using Quilts harmed path finding performance. There are two important differences between these two studies. First, Experiment 1 required use of skip links by constraining the maximum length of "good" paths. When graphs had five layers, this meant that "good" paths contained exactly three links. Experiment 2 did not require using skip links, did not limit maximum path length, and did not have the tight constraint on path length in five-layered graphs. Second, while graphs in Experiment 1 always contained skip links, half the graphs in Experiment 2 had no skip links, meaning that path length was much more directly related to the number of layers. These differences not only varied task difficulties between experiments, but also the amount of practice participants gained in exploiting skip links.

The thresholded effect of nodes mentioned above seems to reappear in three interactions with that variable. First, while increasing link density improved path performance overall, it had a particularly strong impact when the number of nodes was 100. When there were 50 nodes, graphs were quite simple and improved link density had little effect, while when there were 200 nodes, graphs were quite complex and improved link density made little difference. Second, increasing the number of layers reduced performance overall, but had little impact when graphs had 200 nodes. Third, increasing layers improved path finding performance with node-link diagrams, and harmed performance with matrices and Quilts. But both of these effects varied significantly as a function of nodes.

Why were Quilts more effective than the alternative depictions tested here? We believe that Quilts' compactness is an advantage not only in scalability, but also in legibility. Following paths or lines is difficult, but less so when compactness shortens them. This gives Quilts an advantage over both matrices and node-link diagrams. With respect to node-link diagrams, two other issues are worth discussing. First, Quilts do not have node-link diagrams' link crossing problem, making paths easier to follow. Second, unlike node-link diagrams, Quilts do not require distinguishing between oblique lines, something at which the human visual system is particularly weak [4]. Improved path-finding performance with node-link diagrams when the number of layers is large may be applied evidence of this perceptual finding. The same obliqueness in centered matrices (with nodes along a diagonal) may be part of the reason that they support path-finding more poorly than node-link diagrams.

We close our discussion by noting some limitations of our work. Most importantly, to maintain experimental control of the number of layers and node distribution across those layers, we were not able to use layout algorithms for node-link diagrams that freely create and delete layers, and freely reassign nodes to different layers. Node-link depictions may support path-finding better when applications permit exploiting this sort of reassignment, allowing for example skip links to be both shortened and straightened. Our experimental control also precluded the use of real world datasets, which do not easily conform to experimental levels.

## 5 COMPARISON TO PRIOR EXPERIMENTAL WORK

The work we present here is quite unique. To the best of our knowledge, it is the only experimental evaluation of the legibility of *layered* graphs. This evaluation also compares two depictions developed only recently: Quilts and centered matrices. Nevertheless, this work is informed by and has many similarities to previous research.

Ghoniem et al. [8] also compared the legibility of matrices and node-link diagrams for path finding tasks on randomly generated graphs, as link density and the number of nodes varied. While the number of nodes in Ghoniem et al.'s work was set to levels of 20, 50 or 100; we used 50, 100 and 200, in an effort to study legibility of more complex graphs. As we did, Ghoniem et al. found that matrices were slower than node-link diagrams overall, that increasing the number of nodes harmed performance, and that increasing link density improved performance. The relationship they found between depiction and link density was also similar to that we found, with path finding performance using matrices improving rapidly as density increased, and less rapidly (if at all) when node-link diagrams were used. However, the relationship they found between the number of nodes and depiction differs from the relationship we found. In Ghoniem et al., path finding performance with node-link diagrams declined quickly as the number of nodes increased, while performance with matrices remained nearly unchanged. In fact, performance with the two depictions in graphs with 100 nodes was essentially the same. In contrast, we found that path finding performance with both matrices and node-link diagrams increased as the number of nodes increased, and that performance with node-link diagrams was always better than with matrices. It is difficult to explain this difference, given the many possible explanations (e.g. layering, matrix centering, greater complexity).

Keller et al. [12] also compared matrix and node-link legibility for a path finding task as the number of nodes and link density varied, but used graphs drawn from real-life applications to improve external validity. In general, their graphs had fewer links and lower density than ours. They also found an overall advantage for node-link diagrams, that increasing nodes reduced performance, and that increasing links improved performance, especially when matrices were being used.

Henry et al. [9] again compared matrix and node-link legibility, this time for a shortest path finding task, as the number of nodes and link density varied. They also used graphs drawn from real-life applications, which were smaller and less dense than ours. They found an overall advantage for node-link diagrams, and that performance worsened as the number of nodes increased. However, they seem to indicate that increasing link density did not improve performance, as it did in our study and the two others reviewed here.

Overall, whereever our study overlaps with others, we obtained remarkably similar results, giving us confidence in our results.

## 6 CONCLUSION AND FUTURE WORK

Maintaining graph legibility as graph complexity increases is a significant challenge. We addressed this challenge by first improving Quilts, which are more compact than matrix depictions, and do not suffer from the link crossing problems of node-link depictions. We then compared our improved Quilts with node-link diagrams and centered matrices, finding that Quilts always supported the best path finding performance, even as graph complexity increased. Moreover, we presented several novel experimental results describing the legibility of layered graphs, including a decline in path finding performance when skip links were present; and a striking interaction of layers with depiction, with increased layers harming path finding performance when matrix and Quilts depictions are used, and improving performance when node-link diagrams are used.

Future work might compare these depictions when used with un-layered graphs, which permit more freedom in optimizing node-link diagram layout. It would also be very useful to confirm the harmful impact on path finding performance of skip links. Intuitively, one would expect that shortcuts should be a win. Perhaps added layout freedom in node-link diagrams might help realize this win. Should the harmful effect of skip links be confirmed, more research on minimizing them is required: we are aware of only one related work [6], which concerns minimization of dummy nodes rather than skip links. How incompatible are minimizing skip links and layers (or for node-link diagrams, maximizing layers)? Finally, it would be interesting to replicate these results with real-word datasets.

## ACKNOWLEDGMENTS

Quilts were developed with the help of a collaboration with Himesh Patel, Ravinder Devarajan, and David Brink at the SAS Visualization group. At NC State, Theresa-Marie Rhyne of the Center for Visualization and Analytics facilitated the collaboration. Faculty member Matt Stallman made important contributions to quilt design, and helped us with expertise in graph theory. Student Matt Rakow at NCSU developed many early Quilt prototypes. NSF grants 0639426 and 0552802 provided early support for this research.